# Multidecadal forcing of European windstorm losses in CMIP6-DAMIP models

**Stephen Cusack**, Stormwise Ltd, Luton, United Kingdom, stephen.cusack@stormwise.co.uk

**Key Points:**

- Climate models indicate anthropogenic aerosols are likely to have raised European wind losses by tens of percent in the late 20$^{th}$ century
- Further analysis implies major tropical volcanic eruptions contribute significantly to multidecadal windstorm peaks in Europe
- Current lower levels of both anthropogenic and volcanic aerosols provide encouraging signs of lower European windstorm losses in the near future

## Abstract

Large variations of European storm activity at decadal and longer timescales have been found to be driven by major tropical volcanic eruptions, internal climate variability and anthropogenic aerosols (AA). The insurance industry have the ability to align with these slow fluctuations in windstorm risk, yet have not done so, due to uncertainty in how the climate diagnostics used by researchers relate to insured losses. A key aim of this study was to link past research findings to insurance applications by measuring the impacts of AA on European property damage. This was done by extracting the winds from a set of climate model experiments on AA forcing and converting them into European storm losses using an established technique. The multimodel mean result indicates AA boosted insured windstorm losses by 45% in the late 20th century relative to preindustrial times, though model changes range from zero to a doubling of losses and suggest large uncertainty in the size of the strengthening. Further analysis of individual model simulations suggested both AA and internal climate variability explain about half of the amplitude of the last peak, implying a significant role for volcanic eruptions. Looking ahead, we cannot be sure of the timing of the next peak since it may require at least one major volcanic eruption, but climate models provide encouraging signs that recent cuts in AA emissions from North America and Europe will act to reduce its severity.

# 1 Introduction

European windstorm activity has varied at multidecadal scales over the past few centuries (e.g. Dawson et al., 1997; WASA Group, 1998; Brázdil et al., 2004; Mellado-Cano et al., 2019; Hu et al., 2022; Brönnimann et al., 2025). The most recent and best-documented cycle contains particularly large changes in storminess, with average property damages (indexed to 2022) rising from €2.8 Bn in 1960-1979 to €6.7 Bn in 1980-1999, before falling to €2.5 Bn in 2000-2019, based on the storm loss reconstructions in Cusack (2023).

Past research has identified three main drivers of these long-timescale changes in European storm activity. The first driver is major volcanic eruptions injecting huge amounts of sulphur into the tropical stratosphere. Empirical studies of a variety of climate reconstructions suggest this type of volcanic eruption strengthens the European wind climate at multiannual (e.g. Groisman, 1992; Robock and Mao, 1995; Shindell et al., 2004; Fischer et al., 2007; Ortega et al., 2015; Birkel et al., 2018; Brönnimann et al., 2025) and decadal timescales (Dawson et al., 1997; Zanchettin et al., 2013). More detailed analysis provide mechanisms connecting these eruptions to greater storminess (e.g. Robock, 2000; Stenchikov et al., 2009; Mignot et al., 2011). Further, the quasi-periodicity of major tropical volcanic eruptions over the past 600 years (Ammann and Naveau, 2003) has been strongly linked to the phase and timescale of multidecadal variations of climate in the North Atlantic sector (e.g. Otterå et al., 2010; Mann et al., 2021; Dai et al., 2022). However, climate models generally simulate much weaker storm signals from this type of volcanic eruption (e.g. Driscoll et al., 2012; Bittner et al., 2016; Zanchettin et al., 2022) at annual to decadal scales. Considering climate model sensitivity to how the volcanic aerosol is specified (Toohey et al., 2014) and weaknesses in their simulation of stratosphere processes (e.g. Hermanson et al., 2020), particularly the Arctic polar vortex (e.g. Wu and Reichler, 2020), versus the wide variety of climate reconstructions displaying strong volcanic signals, it seems likely that major tropical eruptions are a key driver of European storminess on decadal scales.

Similarly, a large amount of evidence points to internal variability of the climate system driving multidecadal variations of European winds. Climate models consistently produce these slower climate variations in the North Atlantic sector, in the absence of all external forcings (e.g. Zhang et al., 2019; Fang et al., 2021). This driver can be split into three distinct types. The first is is a collection of shorter-timescale anomalies that drive multidecadal variations via unusual extremes, such as the Quasi-Biennial Oscillation (QBO; Baldwin et al., 2001), El Niño-Southern Oscillation (ENSO; Brönnimann, 2007) and the 11-year solar cycle (Gray et al., 2016), and their interactions (Labitzke and Kunze, 2009). The second concerns the Atlantic Meridional Overturning Circulation (AMOC) system of currents flowing through the full depth of the Atlantic Ocean. Variations in the strength of AMOC lead to changes in northward heat transport (e.g. Yan et al., 2018; Robson et al., 2023), which change meridional heat gradients in the northern Atlantic, which in turn modulate storminess in the overlying atmosphere (e.g. Brayshaw et al., 2009; Gastineau and Frankignoul, 2012; Peings and Magnusdottir, 2014; Omrani et al., 2014; Woollings et al., 2015). The mechanisms that cause decadal-scale variations in AMOC are not fully understood, though anomalous atmosphere forcing over the northern Atlantic is considered critical (e.g. Böning et al., 2006; Buckley & Marshall, 2016). The third type of driver is stochastic: the rare chance of weather conditions conducive to extreme events, and the luck of whether storms hit exposure centres to cause larger losses.

Anthropogenic aerosols (AA) are the final main driver of slower variations in the European wind climate. More generally, AA exerted a large influence on temperatures in the North Atlantic (e.g. Booth et al., 2012) and Arctic (e.g. Aizawa et al., 2022) regions in the 20$^{th}$ century, while a variety of recent studies suggest a significant influence of AA on European wind climate. Qin et al. (2020) found significantly steeper meridional pressure gradients over Europe in the late 20th century in the multimodel means from both the fifth and sixth versions of the Coupled Model Intercomparison Project (CMIP5 and CMIP6, e.g. Eyring et al., 2016). The analysis by Hassan et al. (2021) revealed stronger storm track activity in the 1950-1990 period due to AA, and also reported how declines in AA from 1990 to 2020 caused about one-third of the observed slackening of pressure gradients between the subpolar Atlantic and Mediterranean areas, in CMIP6 simulations. Needham and Randall (2023) studied a large ensemble of simulations from a single climate model, and identified changes in AA emissions over North America and Europe as the main driver of observed increases in atmosphere poleward energy transport in northern mid-latitudes over the second half of the 20th century, and subsequent decline this century.

Other external forcings are regarded as minor, relative to the amplitude of multidecadal storm variations. Rising greenhouse gas concentrations are generally considered to produce small changes to near-surface winds and losses in mid-latitude cyclones (e.g. Chang, 2018; Seneviratne et al., 2021; Priestley et al., 2024). Solar radiative forcing anomalies have been relatively small (Gulev et al., 2021) and not aligned with multidecadal cycles of storminess during the industrial period.

The insurance industry has great exposure to the risk of wind damage to property, yet does not apply knowledge of multidecadal storm drivers to their pricing of wind risk. This is mainly because many studies use climate diagnostics with either weak or uncertain relations to the property damage covered by insurance. For example, the North Atlantic Oscillation (NAO; Barnston and Livezey, 1987) is a frequently used guide to European wind climate, and Figure 1 shows a timeseries of its normalized anomalies, alongside windstorm losses. NAO declines through the 1950-70 period differ from the slight rise in storm losses, however, the decoupling of the two in the past decade or so is much more notable. The weak relation to loss fits with findings from other studies on the limited ability of the NAO to describe multidecadal variations in European winter climate (e.g. Woollings et al., 2015; Halifa-Marin et al., 2025). Further, the relation between losses and some other diagnostics, such as maps of anomalies in time-mean surface pressures, seem very reasonable, but not quantified.

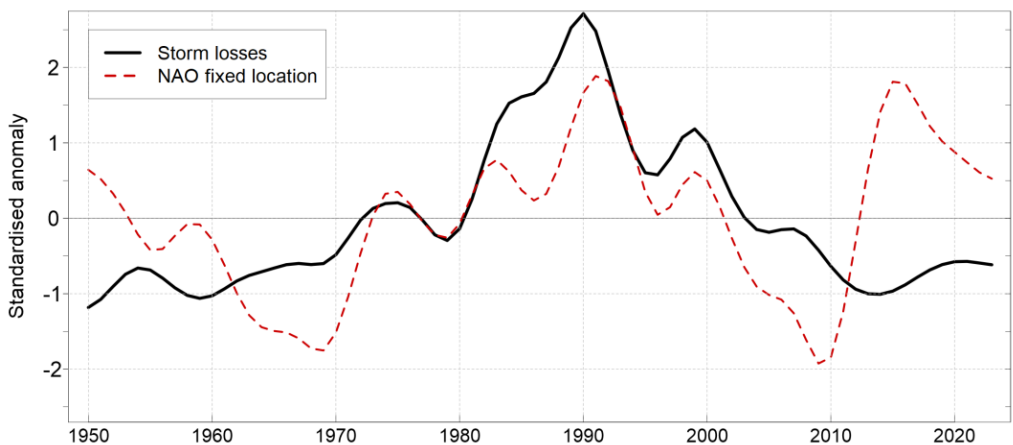

**Figure 1.** Timeseries of windstorm losses and the NAO. Estimates of European windstorm losses are based on an updated version of Cusack (2023), and the station-based NAO index is from University of East Anglia (updated from Jones et al., 1997). All timeseries are low-pass filtered using a second-order Butterworth filter with 10-year cutoff, then values are standardized for inter-comparison.

The main aim of this study is to explore how AA impacted insurance industry losses. Near-surface winds from the DAMIP (Detection and Attribution Model Intercomparison Project) set of climate model experiments (Gillett et al., 2016) were converted to property losses using an established model, then the contribution of AA forcing to variations of European storm damages were estimated. Section 2 contains a description of the DAMIP climate model experiments, and the conversion of their near-surface winds to European storm losses. DAMIP model results on AA-forced storm impacts are presented in Section 3. This is followed by an analysis of its contribution, and that of internal variability, to the last multidecadal peak in the late 20th century. Conclusions are given in Section 4.

**2 Data and methods**

2.1 DAMIP climate model data

CMIP6 includes a series of 23 sub-projects targeting specific questions about the climate. In particular, the DAMIP sub-project was designed to measure the relative roles of various climate forcings during the industrial period. Its Tier 1 experiments consist of climate model simulations covering the 1850-2014 period, with one type of forcing set to historical values and all others fixed at preindustrial values. Initial conditions are from corresponding preindustrial control runs (from the main climate model experiments of CMIP6), and historical forcings are split into three distinct types: natural forcings (solar and volcanic, hereafter Nat), greenhouse gases (GHG), and anthropogenic aerosols (Aero). Participation in CMIP6 sub-projects is optional for climate modeling centres, and this study used climate models from six different modeling centres with the necessary diagnostics for at least five different ensemble members for each Tier 1 test, at the time of download in July 2024. Table 1 summarizes the models and simulations used in this study. Individual simulations are grouped into ensemble means for each of the six models, and the multimodel mean is analysed too: in both cases, the arithmetic mean of the corresponding simulations is used.

**Table 1.** Summary details of DAMIP climate model simulations analysed in this study.

| Model | No. of ensemble members | Control length (yr) | Atmosphere resolution (km) | Ocean resolution (km) |
|---|---|---|---|---|
| CMCC-CM2-SR5 | 8 | 500 | 100 | 100 |
| CanESM5 | 9 | 600 | 500 | 100 |
| HadGEM3-GC31-LL | 15 | 500 | 250 | 100 |
| MIROC6 | 10 | 500 | 250 | 100 |
| MPI-ESM1-2-LR | 15 | 600 | 250 | 250 |
| MRI-ESM2-0 | 5 | 250 | 100 | 100 |

## 2.2 Converting Model Winds to Losses

Windstorm losses in climate simulations were estimated using a method published by Klawa and Ulbrich (2003). Their results were validated by a long timeseries of insurance losses which led to its widespread use (e.g. Leckebusch et al., 2007; Pinto et al., 2007; Karremann et al., 2014; Priestley et al., 2024). A brief overview of their method is now given.

Daily maximum near-surface (10 m) winds in the October to April windstorm season were extracted from model simulations for every grid cell in a study domain consisting of 16 countries in northern and central Europe which experience the vast majority of all insured wind losses. The 98$^{th}$ percentile of the daily maximum wind was computed for each simulation, then a proxy of wind damage ($D_d$) in day d for the whole domain was defined as:

$$D_d = \sum_{i=1}^{N} \left[ max\left(\frac{v_{i,d}}{v_{i,98}} - 1, 0\right) \right]^3$$

where there are N grid cells in the domain, $v_{i,d}$ is the daily maximum wind for the i'th grid cell on day d, and $v_{i,98}$ is the climatological 98$^{th}$ percentile of wind for the i'th grid cell. Storm events (s) are then defined as a series of up to three days centred on the days with peak values of $D_d$, and the event maximum wind ($v_{i,s}$) is set to the maximum of the daily values comprising the storm, for each grid cell.

Finally, domain-wide event losses were estimated from the storm-maximum winds:

$$L_s = c . \sum_{i=1}^{N} P_i \left[ max\left(\frac{v_{i,s}}{v_{i,98}} - 1, 0\right) \right]^3$$

where $L_s$ is the loss for storm s, and $P_i$ is the population count for the i'th cell from Gridded Population of the World, version 4, at 2.5 minutes of arc resolution (CIESIN, 2018). The constant of proportionality (c) in the above loss equation was used by KU03 to re-scale their index values to actual loss data. Results in section 3 concern relative change in losses, hence setting this coefficient to unity does not alter findings.

## 3 Results and Discussion

Figure 2a shows the timeseries of multi-model mean change in Europe-wide storm losses due to AA forcing, based on all 62 DAMIP simulations. A low-frequency second-order Butterworth filter has been applied to modeled losses, using a 20-year cutoff to highlight multidecadal variations, while reducing shorter timescale climate variations such as the 11-year solar cycle and ENSO. The Aero ensemble has a peak at the end of the 20th century with 45% greater losses than its preindustrial control run, and more than four standard errors above zero.

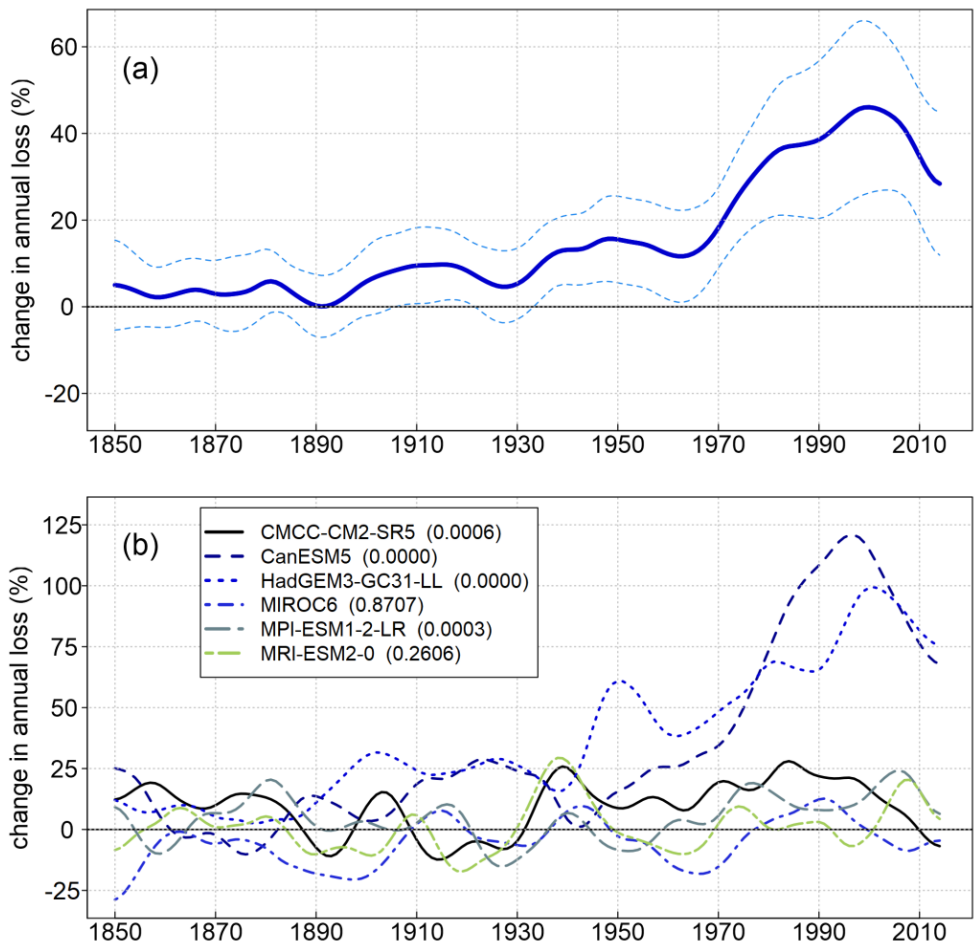

**Figure 2.** AA-forced change in modeled losses. (a) Timeseries of change (%) in DAMIP multimodel mean Europe-wide storm losses due to AA forcing, together with uncertainty (two standard errors of the mean, dashed lines). (b) Ensemble-mean AA forcing of European windstorm losses in each of the six climate models. The p-value of a null hypothesis that the model has zero anomaly in the 1970-2009 period is given in the legend.

Analysis of individual model ensemble-means reveal the simulated signal is less certain, as shown in Figure 2b. A t-test was performed on the null hypothesis that each model's mean anomaly was zero, using the longer 1970-2009 period for more robust results. The p-values of the null hypothesis for each model, given in parentheses in the legend of Figure 2b, indicate four models with a signal above zero at the 99.9% confidence level, with two models projecting a near-certain increase in storm loss due to AA. However, two models simulate no significant AA forcing on European storm losses.

There is evidence that the AA forcing in CMIP6 models may be too strong. A portion of these models lie outside the range of observed values for (i) shortwave radiative forcing trends since 1985 (Menary et al., 2020) and (ii) observed surface air temperature trends over North America (Robson et al., 2022), both of which were linked by the authors to strong indirect aerosol forcing. Though models tend to underestimate atmosphere responses in the northern Atlantic to various forcings (e.g. Scaife et al., 2014; Eade et al., 2014; Weisheimer et al., 2024), hence biases in AA-simulated storm impacts depend on other model aspects, beside the strength of AA forcing. This is illustrated using information in Robson et al. (2022) on the strength of AA-forcing for various models, which includes four of the six studied here. Notably, MRI-ESM2-0 has strong AA forcing, yet is one of the two models with no storm impacts (Figure 2b), while MPI-ESM1-2-LR has the weakest AA forcing of their 17 studied models yet it causes significant storm impacts (Figure 2b). Further research into both the magnitude of AA forcing and how it modulates storminess could reduce uncertainty in modeling results seen in Figure 2b.

The AA forcing of storminess is accompanied by large-scale circulation anomalies. Figure 3 shows a map of the AA-forced geopotential height anomaly in the 1970-2009 period, relative to preindustrial control simulations, for the DAMIP multimodel mean. The meridional gradients of heights in mid-latitudes over the North Atlantic sector correspond to anomalous westerlies over Europe, and indicate AA forcing drives anomalies in the large-scale atmospheric circulation which in turn increase windstorm losses.

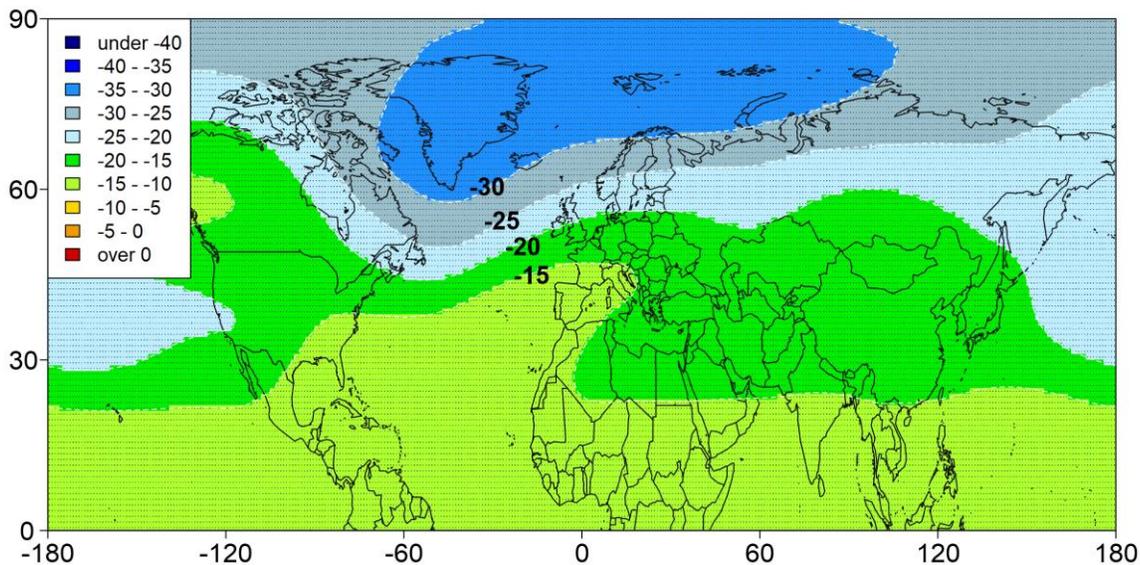

**Figure 3.** AA-forced changes in geopotential height at 500 hPa. DAMIP multimodel-mean geopotential height anomalies at 500 hPa (m) for the 1970-2009 period relative to preindustrial control runs. Areas with anomalies more than two standard errors from zero are stippled.

The results so far concern the AA boost to storminess from the preindustrial period (1850 CE forcing levels) to the late 20th century. We now shift the focus to understanding the observed change in storminess over the past few decades in the remainder of this section.

From Figure 1, the most recent multidecadal cycle consists of lower losses in the 1950s and 1960s, rising to a peak in the late 20th century, then a subsequent decline. DAMIP model simulations with AA forcing have average storm losses in 1980-1999 which were 33% higher than in 1950-1969 (different from zero at the 1% level). In contrast, the reconstruction of historical storm losses in Cusack (2023) suggest average storm losses in 1980-1999 were 242% higher than in 1950-1969. Therefore, the models indicate AA forcing explains a small amount of the steep rise in losses during the second half of the 20th century.

The analysis of individual ensemble members of DAMIP can provide new information on the recent multidecadal cycle. Each simulation contains both AA forcing, and an independent realization of internal climate variability. We now examine the amplitude of multidecadal variations of loss in each of the 62 ensemble members, to assess whether the combined effects of AA external forcing and natural internal variability can explain the amplitude of multidecadal change observed in recent decades. First, the portion from 1940 to 2014 of each model simulation was extracted, to obtain 62 individual segments. Second, rolling 20-year mean losses were computed in each segment. Finally, the maximum value of 20-year mean loss was expressed as a fraction of the average loss in that segment. The distribution of these values in the 62 segments is shown in Figure 4, together with the corresponding observed value, based on the mean 1980-1999 loss divided by the 1940-2014 average loss.

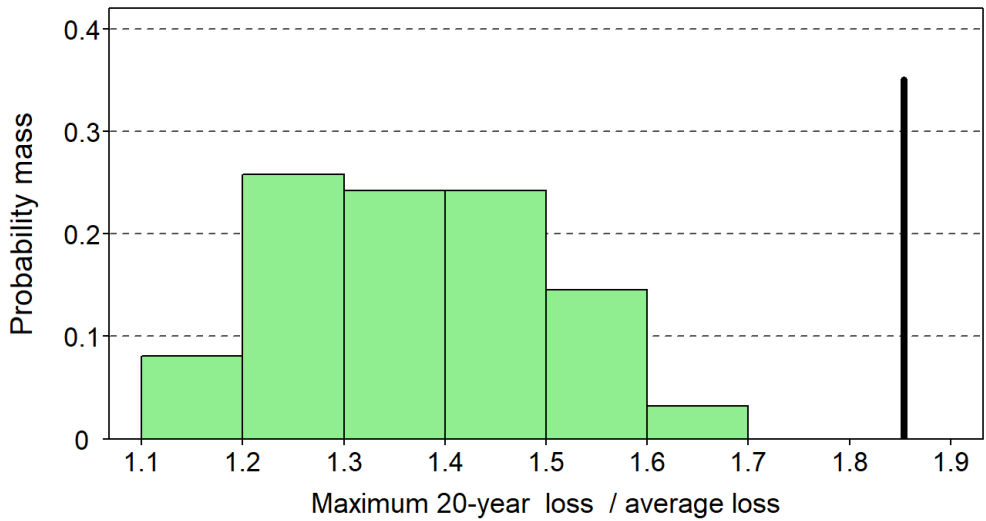

**Figure 4.** Size of simulated multidecadal peaks, versus observed. The maximum 20-year loss as a fraction of the average loss in the 1940-2014 period was computed for each of the 62 ensemble members of DAMIP AA-forced simulations. The distribution of these relative peaks is plotted. The vertical black line shows the corresponding estimate from a wind loss reconstruction in 1940-2014 (updated from Cusack, 2023).

Results in Figure 4 reveal the combination of internal and AA forcings drives multidecadal peaks which are typically 25% to 65% of the observed amplitude in the 1980-99 period. While this is material to those with exposure to storm losses, it is also notable how, on average, the modeled peak is 43.8% of the amplitude of the late 20$^{th}$ century peak, and the ensemble member with the greatest relative peak is only 74.6% of what occurred.

Other studies also find climate models tend to underestimate multidecadal climate variability (e.g. Wang et al., 2017; Kim et al., 2018). Modeling experiments by Kim et al. (2018) found coupled climate models simulate a realistic ocean response to atmosphere forcing, but do not capture the full extent of observed ocean and sea-ice forcing of the atmosphere. This suggests weak coupling from the lower boundary to the atmosphere contributes to the low bias in modeled estimates of multidecadal variability seen in Figure 4. Though the size of this contribution seems small: Kim et al. (2018) found subpolar Atlantic temperature anomalies linearly explain about 10% of the multidecadal variance of the NAO, when the former leads the latter. Therefore, about half of the observed multidecadal peak in 1980-1999 remains unexplained by the DAMIP simulations with internal and AA forcings, after allowing for model bias in the simulated ocean forcing of the atmosphere.

Kim et al. (2018) concluded that the shortfall in modelled multidecadal variability is likely due to underestimated wind anomalies in the atmosphere model. We suggest the unexplained 50% could be largely explained by the exclusion of volcanic eruptions from the DAMIP simulations used in Figure 4. A separate DAMIP experiment ("hist-nat") was designed to measure this effect (and solar variations too), however, as discussed in the Introduction, climate models have documented shortcomings in their simulation of volcanic impacts on mid-latitude winter dynamics, and we found very little volcanic impacts on storminess in hist-nat (not shown). The Introduction also mentioned the evidence from multiple independent climate reconstructions which strongly suggest volcanic eruptions drive multiannual variations in atmosphere circulation, and other studies suggesting they influence the phase and amplitude of past multidecadal variations. Further, the late 20$^{th}$ century peak coincided with the rare occurrence of two major volcanic eruptions inside 10 years (1982 and 1991). Based on these considerations, we propose volcanic eruptions caused most of the unexplained half of the multidecadal variation in 1980-1999. A better understanding of why models produce much weaker volcanic signals than observed could reduce uncertainty in the relative roles of multidecadal storm drivers.

**4 Conclusions**

CMIP6-DAMIP multimodel simulations indicated AA boosted European windstorm losses by 45% in the late 20th century relative to preindustrial times. While all six models indicate AA drives increases in storminess, there is large uncertainty in the size of the storm impacts, since signals from individual climate model ensembles vary from near-zero to 100%. Further research into the size of AA forcing and how it modulates storminess would help reduce uncertainty.

The AA contribution to the large multidecadal variation of European windstorm losses since the mid-20$^{th}$ century was found to be minor. DAMIP simulations with AA forcing had 33% higher storm losses in 1980-1999 compared to 1950-1969. While this change is significant at the 1% level, it explains only a small part of the 242% increase from a historical reconstruction of storm losses.

Individual DAMIP simulations contain both internal variability and AA forcings, and further analysis revealed these two drivers explain about half of the amplitude of the peak loss anomaly in 1980-1999, after accounting for model underestimates of ocean forcing of the atmosphere. We suggest that the other known multidecadal driver, namely major, tropical, sulphur-rich volcanic eruptions, contributed most of the other half to peak storminess in the late 20$^{th}$ century. This inference fits with past studies reporting a main role for volcanic forcing of multidecadal climate, and with the occurrence of two major climate-changing volcanoes in 1982 and 1991. However, climate model simulations of major volcanic eruptions generally produce much smaller storm changes than observed. A better understanding of why models contain such a small volcanic signal, and their more general underestimation of mid-latitude dynamical signals at seasonal and longer timescales, would increase our confidence in the relative contributions of drivers toward multidecadal storminess in Europe.

While uncertainties remain substantial, past research provides information that could improve our view of windstorm risk. For example, past multidecadal storm fluctuations have a roughly 75-year duration, and views of wind climate based on shorter periods may be biased. Further, how likely is the occurrence of the full strength of the late 20th century storm peak in the near future, given reduced emissions of anthropogenic aerosols from North America and Europe, and the absence of major, tropical, sulphur-rich volcanoes? Consideration of these issues could lead to more accurate views of windstorm risk.

**Acknowledgments**

Many thanks to Tyler Cox and Matt Raywood for their feedback on earlier versions of this manuscript.